\documentclass[seceq]{ptptex}

\usepackage{graphicx}
\usepackage{wrapft}

\title{
An analytical approximation of the luminosity distance \\ in flat
  cosmologies with a cosmological constant
}

\author{
Masaru \textsc{Adachi}
and Masumi \textsc{Kasai}\footnote{E-mail: kasai@phys.hirosaki-u.ac.jp}
}

\inst{
Graduate School of Science and Technology, Hirosaki University,\\
Hirosaki 036-8561, Japan
}

\abst{ We present an analytical approximation formula for the
  luminosity distance in spatially flat cosmologies with dust and a
  cosmological constant.  Apart from the overall factor, the effect of
  non-zero cosmological constant in our  formula is written
  simply in terms of a rational function.  We also show the
  approximate formulae for the Dyer-Roeder distance (empty beam case)
  and the generalized angular diameter distance from redshift $z_1$ to
  $z_2$, which are particularly useful in analyzing the gravitational
  lens effects.  Our formulae are widely applicable over the range of
  the density parameter and the redshift with sufficiently small
  uncertainties.
  In particular, in the range of density parameter $0.3 \leq
  \Omega_{\rm m} \leq 1$ and redshift $0.03 \leq z \leq 1000$, the
  relative error for the luminosity distance by our formula is always
  smaller than that of the recent work by Wickramasinghe and Ukwatta
  (2010).  Hence, we hope that our formulae will be an efficient and useful tool
  for exploring various problems in
  observational cosmology.  }

\begin{document}

\maketitle

\section{Introduction}

Current cosmological observations indicate that the universe is
spatially flat with a cosmological constant. 
In such a universe, calculations of the luminosity distance and the
angular diameter distance require numerical integrations and elliptic
functions \cite{eisen}.  
In order to simplify the repeated numerical integrations, 
Pen\cite{pen} has developed an efficient fitting formula. 
Recently, Wickramasinghe and Ukwatta\cite{wu} have shown another analytical method, which runs
faster than that of Pen\cite{pen} and has smaller error variations with
respect to redshift $z$. (See also Ref.~\citen{liu}.)

In this paper, we present yet another analytical approximation to
calculate the luminosity distance as follows:
\begin{equation}\label{eq:one}
  d_{\rm{L}}(z,\Omega_{\rm m}) = \frac{2c}{H_0}
  \frac{1+z}{\sqrt{\Omega_{\rm m}}}
    \left\{
    \Phi\big(x(0, \Omega_{\rm m})\big)
    - \frac{1}{\sqrt{1+z}} \Phi(x(z,\Omega_{\rm m}))\right\},
\end{equation}
\begin{equation}\label{eq:two}
  \Phi(x) = 
  \frac{1 + 1.320 x + 0.4415 x^2 + 0.02656 x^3}
       {1 + 1.392 x + 0.5121 x^2 + 0.03944 x^3},
\end{equation}
where $c$ is the speed of light, $H_0$ is the Hubble constant,
$\Omega_{\rm m}$ is the density parameter of dust matter, related
to the density parameter of vacuum energy $\Omega_{\Lambda}$ by
$\Omega_{\rm m}+\Omega_{\Lambda}=1$, and
\begin{equation}\label{eq:three}
  x(z,\Omega_{\rm m})
  = \frac{1-\Omega_{\rm m}}{\Omega_{\rm m}}  \frac{1}{(1+z)^3}. 
\end{equation}
Apart from the overall factor $1/\sqrt{\Omega_{\rm m}}$, 
the effect of non-zero cosmological constant in
our distance formula is written simply in terms of a rational
function $\Phi(x)$. 

The function $\Phi(x)$ has the following properties:  
\begin{enumerate}
\item $\Phi = 1$ for $x=0$.
\item $\Phi$ is a monotonically decreasing function of $x$,  
  $d\Phi/dx<0$, and $\Phi\rightarrow 0.6735$ for
  $x\rightarrow \infty$. 
\item $\Phi$ is a monotonically increasing function of
  $\Omega_{\rm m}$,  $\partial \Phi/\partial
  \Omega_{\rm m}>0$, and
  $\Phi\rightarrow 1$ for $\Omega_{\rm m}\rightarrow 1$.
\item $\Phi$ is a monotonically increasing function of $z$,  
  $\partial \Phi/\partial z>0$, and  $\Phi\rightarrow 1$
  for $z\rightarrow \infty$.
\item $0.6735 < \Phi(x(0,\Omega_{\rm m})) < \Phi(x(z,\Omega_{\rm
    m})) < 1$ for $0<z$, $0 < \Omega_{\rm m} < 1$. 
\end{enumerate}
Note that our approximate formula is explicitly shown to be exact
when $\Omega_{\rm m}=1$: 
\begin{equation}
  d_{\rm L}(z,1)=\frac{2c}{H_0}
  (1+z) 
    \left\{
    1
    - \frac{1}{\sqrt{1+z}}\right\}. 
\end{equation}

\section{Approximation}

The luminosity distance in flat cosmologies with a cosmological
constant is given by
\begin{equation}
  d_{\rm{L}}(z,\Omega_{\rm m}) = \frac{c}{H_0} (1+z) \int_{\frac{1}{1+z}}^1
  \frac{da}{\sqrt{\Omega_{\rm{m}} a + (1-\Omega_{\rm{m}}) a^4}}.
\end{equation}
We define 
\begin{equation}
  F =  \int_0^a
  \frac{\sqrt{\Omega_{\rm{m}}} \,da'}{\sqrt{\Omega_{\rm{m}} a' + (1-\Omega_{\rm{m}}) a'^4}}. 
\end{equation}
The power series expansion of $F$ with respect to $a$ around $a=0$
yields 
\begin{equation}
  F = \sqrt{a} \left(2 - \frac{1}{7} x
    + \frac{3}{52} x^2
    - \frac{5}{152} x^3
    + \cdots \right), 
\end{equation}
where
\begin{equation}
  x = \frac{1-\Omega_{\rm{m}}}{\Omega_{\rm{m}}} a^3. 
\end{equation}
After expanding $F$ up to $O(x^6)$, we can obtain 
 the Pad\'e approximant to the following order:
 \begin{equation}
   F = \sqrt{a} \frac{2 + b_1 x + b_2 x^2 + b_3 x^3}
   {1 + c_1 x + c_2 x^2 + c_3 x^3}, 
 \end{equation}
where the numerical constants are determined as follows:
\begin{equation}
  b_1 = \frac{4222975319}{1599088274}, 
\end{equation}
\begin{equation}
  b_2 = \frac{1138125153117}{1288865148844}, 
\end{equation}
\begin{equation}
  b_3 = \frac{7433983569773}{139933930445920}, 
\end{equation}
\begin{equation}
  c_1 = \frac{635916643}{456882364}, 
\end{equation}
\begin{equation}
  c_2 = \frac{14505955555}{28326706568}, 
\end{equation}
\begin{equation}
  c_3 = \frac{44686179629}{1133068262720}. 
\end{equation}
Setting $a=1/(1+z)$, we finally obtain
Eqs.~(\ref{eq:one})--(\ref{eq:three}).

In order to compare our method with that of Ref.~\citen{wu}, we calculate
the following relative error  
\begin{equation}
  \Delta E = \frac{|d_{\rm{L}}^{\rm{appr}} - d_{\rm{L}}^{\rm{num}}|}
  {d_{\rm{L}}^{\rm{num}}}\times 100 \ (\rm{per\ cent}), 
\end{equation}
where $d_{\rm{L}}^{\rm{appr}}$ and $d_{\rm{L}}^{\rm{num}}$ represent the values of
luminosity distances calculated by using approximate formula and
numerical method, respectively. 

\begin{wrapfigure}{l}{6.6cm}
\includegraphics[width=6.6cm]{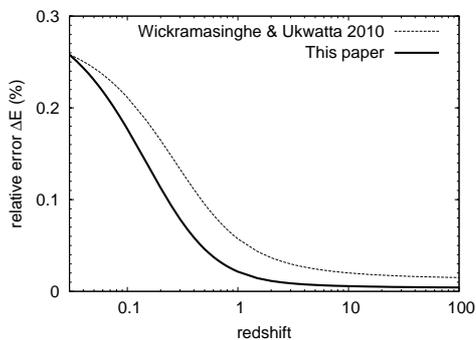}
\caption{A comparison of the percentage relative error $\Delta E$ for two
  analytical methods as a function of the redshift for
  $\Omega_{\rm{m}}=0.3$.}
\label{fig:one}
\end{wrapfigure}
Fig.~\ref{fig:one} shows a comparison of $\Delta E$ for both analytical
methods for $\Omega_{\rm{m}} = 0.3$. 
It is shown that our method has
a smaller relative error for redshift range $0.03 \leq z \leq
100$. Although the relative error for our method is slightly worse in
the range $0.01 \leq z < 0.03$, it is still less than $0.3$ per cent.

Since our method is based on the Taylor expansion and the Pad\'e
approximant with respect to $x=a^3 (1-\Omega_{\rm{m}})/\Omega_{\rm{m}}$, it is
evident that the error in our method decreases monotonically with
increasing redshift $z$ (or increasing $\Omega_{\rm{m}}$), i.e., with
decreasing $x$.

\begin{table}
\caption{The percentage relative error $\Delta E (\%)$ for the
  luminosity distance by our formula Eq.~(\ref{eq:one})} 
\label{table:one}
\begin{center}
\begin{tabular}{lrrrrr}
\hline\hline
$\Omega_{\rm{m}}$ & $z=0.03$ & $z=0.1$ & $z=1$ & $z=10$ & $z=1000$ \\
\hline
$0.2$ &  1.87\% & 1.38\% & 0.25\% & 0.08\% & 0.06\% \\   
$0.3$ &  0.26\% & 0.18\% & 0.02\% & 0.01\% & $<0.01\%$ \\   
$0.4$ &  0.03\% & 0.01\% & $<0.01\%$ & $<$0.01\% & $<0.01\%$ \\   
$1$ &  0 & 0 & 0 & 0 & 0 \\  
\hline 
\end{tabular}
\end{center}
\end{table}

Table \ref{table:one} shows the relative percentage error $\Delta E$
in our method. It is apparent that our approximate formula has 
sufficiently small uncertainties in the wide range of parameters. 
Only one exception is the nearby region (say, $z< 0.1$) in the low
density (say, $\Omega_{\rm{m}} < 0.2$) universe, where the relative error
$\Delta E$ exceeds $1$ per cent. 
In such a nearby region ($z\ll 1$), however, we may alternatively use
the power series expansion around $z=0$ with sufficient accuracy. 

In particular, in the range of density parameter $0.3 \leq \Omega_{\rm
  m} \leq 1$ and redshift $0.03 \leq z \leq 1000$, the relative error
for the luminosity distance in our formula is always smaller than that
of the recent work by Wickramasinghe and Ukwatta\cite{wu}.

\section{The empty beam case}

The distance formula which takes the effect of clumpy distribution of
matter into account has proposed in Refs.~\citen{zel,da}, and later by
Dyer and Roeder\cite{dr} in more general form, which is now known as
the Dyer-Roeder distance. The Dyer-Roeder luminosity distance for
empty beam case in flat cosmologies is
\begin{equation}
  D_{\rm{L}}(z,\Omega_{\rm m}) = \frac{c}{H_0}(1+z)^2\int_{\frac{1}{1+z}}^1
  \frac{a^2\,da}{\sqrt{\Omega_{\rm{m}} a + (1-\Omega_{\rm{m}})a^4}}. 
\end{equation}
Our analytic formula for the empty beam is
\begin{equation}\label{eq:DLG}
D_{\rm{L}}(z,\Omega_{\rm m}) =
\frac{2}{5}\frac{c}{H_0}\frac{(1+z)^2}{\sqrt{\Omega_{\rm m}}}
\left\{ \Psi(x(0,\Omega_{\rm m}))
  - \frac{1}{(1+z)^{\frac{5}{2}}}\Psi(x(z,\Omega_{\rm m}))\right\}, 
\end{equation}
\begin{equation}\label{eq:G}
  \Psi(x) = 
  \frac{1 + 1.256 x + 0.3804 x^2 + 0.0164 x^3}{1 + 1.483x +
 0.6072 x^2 + 0.0587x^3}, 
\end{equation}
where $x$ is defined in Eq.~(\ref{eq:three}). The formula can be
obtained in the same way as that in the previous section. The rational
function $\Psi(x)$ has the following properties:
\begin{enumerate}
\item $\Psi = 1$ for $x=0$.
\item $\Psi$ is a monotonically decreasing function of $x$,  
  $d\Psi/dx<0$, and $\Psi\rightarrow 0.2792$ for
  $x\rightarrow \infty$. 
\item $\Psi$ is a monotonically increasing function of
  $\Omega_{\rm m}$,  $\partial \Psi/\partial
  \Omega_{\rm m}>0$, and
  $\Psi\rightarrow 1$ for $\Omega_{\rm m}\rightarrow 1$.
\item $\Psi$ is a monotonically increasing function of $z$,  
  $\partial \Psi/\partial z>0$, and  $\Psi\rightarrow 1$
  for $z\rightarrow \infty$. 
\item $0.2792 < \Psi(x(0,\Omega_{\rm m})) < \Psi(x(z,\Omega_{\rm
    m})) < 1$ for $0<z$, $0 < \Omega_{\rm m} < 1$. 
\end{enumerate}
Again, our approximate formula Eq.~(\ref{eq:DLG}) is exact when
$\Omega_{\rm m}=1$:
\begin{equation}
  D_{\rm{L}}(z,1) =
  \frac{2}{5}\frac{c}{H_0}(1+z)^2
\left\{ 1
  - \frac{1}{(1+z)^{\frac{5}{2}}}\right\}. 
\end{equation}
Note that no consideration on the Dyer-Roeder distance has been taken
in the previous papers, e.g., Refs.~\citen{pen,wu}.

\begin{wraptable}{l}{\halftext}
\caption{The percentage relative error $\Delta E(\%)$ for the
Dyer-Roeder distance (empty beam case) by our formula 
 Eq.~(\ref{eq:DLG})}  
\label{table:two}
\begin{center}
\begin{tabular}{lrrrr}
\hline\hline
$\Omega_{\rm{m}}$ & $z=0.03$ & $z=0.1$ & $z=1$ & $z=10$  \\
\hline
$0.2$ & 0.90\% & 0.68\% & 0.21\% & 0.15\%  \\   
$0.3$ & 0.14\% & 0.10\% & 0.03\% & 0.02\%  \\   
$0.4$ & 0.02\% & 0.01\% & $<0.01\%$ & $<0.01\%$  \\   
$1$ &  0 & 0 & 0 & 0  \\  
\hline 
\end{tabular}
\end{center}
\end{wraptable}
Table~\ref{table:two} shows the relative error of the our formula
Eq.~(\ref{eq:DLG}).  The relative error for $0.3 \leq \Omega_{\rm{m}}
\leq 1$ is always less than $0.15$ per cent in the range $0.03 \leq z
\leq 10$.  For $\Omega_{\rm{m}}=0.2$, the accuracy gets slightly
worse, but the error is still less than $1$ per cent in the same
redshift range.

We omitted the error calculations for $z > 10$ for the following
reasons. First, the errors are sufficiently small in those regions,
and second, the Dyer-Roeder description is relevant only in the
regions where the clumpy distribution of matter becomes important.

\section{The generalized angular diameter distance}

Here we present the analytic formulae for the generalized angular
diameter distance from redshift $z=z_1$ to $z=z_2$, which is
frequently required in analyzing the gravitational lens effects.  Here
we only consider the case of flat cosmologies. For the distance
formulae in non-flat cases, see, e.g., Ref.~\citen{ffkt}.  For the
standard (filled beam) case, the angular diameter distance from
redshift $z=z_1$ to $z=z_2$ is
\begin{equation}
  d_{\rm A}(z_1, z_2) = \frac{c}{H_0}\frac{1}{1+z_2}
  \int_{\frac{1}{1+z_2}}^{\frac{1}{1+z_1}}
  \frac{da}{\sqrt{\Omega_{\rm{m}} a + (1-\Omega_{\rm{m}}) a^4}}.
\end{equation}
Our analytical approximation is simply 
\begin{equation}\label{eq:dak20} 
  d_{\rm{A}}(z_1, z_2) = \frac{2c}{H_0}\frac{1}{\sqrt{\Omega_{\rm{m}}}(1+z_2)}
\left\{
  \frac{1}{\sqrt{1+z_1}}\Phi(x(z_1, \Omega_{\rm m}))
  - \frac{1}{\sqrt{1+z_2}}\Phi(x(z_2, \Omega_{\rm m}))
\right\}, 
\end{equation}
where $\Phi(x)$ is defined in Eq.~(\ref{eq:two}).  The relative error
of our formula Eq.~(\ref{eq:dak20}) for $\Omega_{\rm{m}}=0.3$ is less
than $0.02$ per cent in the range $0 \leq z_1 < 1$ for $z_2=1$, and
less than $0.01$ per cent in the range $0 \leq z_1 < 3$ for $z_2=3$.

For the empty beam case, 
\begin{equation}
  D_{\rm{A}}(z_1, z_2)  = \frac{c}{H_0}(1+z_1)
  \int_{\frac{1}{1+z_2}}^{\frac{1}{1+z_1}}
  \frac{a^2\,da}{\sqrt{\Omega_{\rm{m}} a + (1-\Omega_{\rm{m}}) a^4}}, 
\end{equation}
and our approximate formula is
\begin{equation}
  D_{\rm{A}}(z_1, z_2)  = \frac{2}{5}
  \frac{c}{H_0}\frac{(1+z_1)}{\sqrt{\Omega_{\rm{m}}}}
  \left\{
    \frac{1}{(1+z_1)^{\frac{5}{2}}} \Psi(x(z_1,\Omega_{\rm{m}})) 
- \frac{1}{(1+z_2)^{\frac{5}{2}}}\Psi(x(z_2,\Omega_{\rm{m}}))\right\},
\end{equation}
where $\Psi(x)$ is
already defined in Eq.~(\ref{eq:G}). 

The reciprocity theorem holds between the angular diameter
and luminosity distances as follows:
$d_{\rm{L}}(z) = (1+z)^2 d_{\rm{A}}(0, z)$, and 
$D_{\rm{L}}(z) = (1+z)^2 D_{\rm{A}}(0,z)$.

\section{Summary}

We have presented a simple analytical approximation formula for the
luminosity distance in flat cosmologies with a cosmological constant.
We have also shown the approximate formulae for the Dyer-Roeder
distance and the generalized angular diameter distance from redshift
$z=z_1$ to $z=z_2$, which are particularly useful in analyzing the
gravitational lens effects.  Apart from the overall factor
$1/\sqrt{\Omega_{\rm m}}$, the effects of non-zero cosmological
constant in our distance formulae are written simply in terms of the
rational functions $\Phi(x)$ for ``filled beam case'' and $\Psi(x)$
for ``empty beam case''. Both are monotonically decreasing functions
with respect to $x$, and increasing ones with respect to redshift $z$
and the density parameter $\Omega_{\rm m}$.

Our formulae are widely applicable over the range of the density
parameter and the redshift with sufficiently small uncertainties. 
In particular, in the range 
$0.3 \leq \Omega_{\rm m} \leq 1$ and  $0.03 \leq z \leq 1000$, 
the relative error for the luminosity distance by our formula is
always smaller than that of the recent work by Wickramasinghe and
Ukwatta\cite{wu}. 
Hence, we hope that it will be an efficient and useful tool for
exploring various problems in observational
cosmology, such as the statistical grativational lensing,  the
cosmological parameter fitting in the magnitude-redshift relation of
the supernovae, and so on.

\appendix
\section{Approximations in a small redshift region}

In the region of small redshift ($z \ll 1$), the standard power series
expansions can safely be used.  The power series expansions of
$d_{\rm{L}}(z,\Omega_{\rm m})$ (filled beam case) and
$D_{\rm{L}}(z,\Omega_{\rm m})$ (empty beam case) around $z=0$ are
\begin{equation}\label{a1}
  d_{\rm{L}}(z,\Omega_{\rm m})=\frac{c}{H_0}  \left\{
    z +\left(1- \frac{3}{4}\Omega_{\rm{m}}\right) z^2
    + \frac{9\Omega_{\rm{m}}-10}{8}\Omega_{\rm{m}} z^3 + \cdots
\right\},
\end{equation}
\begin{equation}\label{a2}
  D_{\rm{L}}(z,\Omega_{\rm m})=\frac{c}{H_0} \left\{
z + \left(1- \frac{3}{4}\Omega_{\rm{m}}\right) z^2
+ \left(\frac{9}{8}\Omega_{\rm{m}} - 1\right) \Omega_{\rm{m}}z^3 + \cdots
\right\}. 
\end{equation}
Just for reference, their Pad\'e approximants are
\begin{equation}\label{a3}
  d_{\rm{L}}(z,\Omega_{\rm m}) = \frac{c}{H_0} \frac{(12\Omega_{\rm{m}}-16)z +
    (9\Omega_{\rm{m}}^2+4\Omega_{\rm{m}}-16)z^2}
  {(12\Omega_{\rm{m}}-16) + (18\Omega_{\rm{m}}^2-20\Omega_{\rm{m}})z}, 
\end{equation}
\begin{equation}\label{a4}
  D_{\rm{L}}(z,\Omega_{\rm m})=\frac{c}{H_0} \frac{(12\Omega_{\rm{m}}-16)z +
    (9\Omega_{\rm{m}}^2+8\Omega_{\rm{m}}-16)z^2}
  {(12\Omega_{\rm{m}}-16) + (18\Omega_{\rm{m}}^2-16\Omega_{\rm{m}})z}. 
\end{equation}

\begin{table}
\caption{The percentage relative error $\Delta E(\%)$ of the
power series expansion Eq.~(\ref{a1})}  
\label{table:a1}
\begin{center}
\begin{tabular}{lrrrrr}
\hline
$\Omega_{\rm{m}}$ & $z=0.01$ & $z=0.1$ & $z=0.2$ & $z=0.5$& $z=1$  \\
\hline
$0.2$ & $<0.01\%$ & $<0.01\%$ & $0.02\%$ & $0.10\%$ & $0.12\%$\\   
$0.5$ & $<0.01\%$ & $0.01\%$& $0.09\%$ & $1.25\%$ & $8.04\%$\\   
$0.7$ & $<0.01\%$ & $0.02\%$& $0.15\%$ & $1.80\%$ & $10.56\%$\\   
$1.0$ & $<0.01\%$ & $<0.01\%$ & $0.05\%$ & $0.66\%$ &  $3.98\%$\\  
\hline 
\end{tabular}
\end{center}
\end{table}
\begin{table}
\caption{The percentage relative  error  $\Delta E(\%)$ of the 
 Pad\'e approximant  
 Eq.~(\ref{a3})}  
\label{table:a2}
\begin{center}
\begin{tabular}{lrrrrr}
\hline
$\Omega_{\rm{m}}$ & $z=0.01$ & $z=0.1$ & $z=0.2$ & $z=0.5$& $z=1$  \\
\hline
$0.2$ & $<0.01\%$ &$<0.01\%$ & $0.05\%$ & $0.51\%$ & $2.30\%$\\   
$0.5$ & $<0.01\%$ &$<0.01\%$ & $0.03\%$ & $0.24\%$ & $0.72\%$\\   
$0.7$ & $<0.01\%$ &$<0.01\%$ & $<0.01\%$& $0.05\%$ & $0.36\%$\\   
$1.0$ & $<0.01\%$ &$<0.01\%$ & $<0.01\%$& $0.09\%$ & $0.42\%$\\  
\hline 
\end{tabular}
\end{center}
\end{table}
The relative error of Eqs.~(\ref{a1}) and (\ref{a3}) are listed in
Tables~\ref{table:a1} and \ref{table:a2}.  The maximal relative error
in the power series expansion $\Delta E$ is $10.6$ per cent at $z=1$
in the interval $0 < z < 1$ and $0.2 \leq \Omega_{\rm{m}} \leq 1.0$,
not $37$ per cent which was claimed by Pen\cite{pen}.  The Pad\'e
approximant Eq.~(\ref{a3}) shows in many cases better accuracy than
the power series expansion Eq.~(\ref{a1}). The relative error of the
Pad\'e approximant does not exceed $3$ per cent even at redshift $z=1$
in the rage $0.2 \leq \Omega_{\rm{m}}\leq 1.0$.

\section{An analytical approximation of the growth function}

There is one more thing.  Here we present another approximate formula which
we hope to be an efficient and useful tool in observational cosmology.  
Heath\cite{heath} has shown that the
growth function in a dust cosmology can be written as
\begin{equation}
  D_1(a) \propto H(a) \int_0^a \frac{da'}{\big(a' H(a')\big)^3},
\end{equation}
\begin{equation}
  D_2 \propto H(a), 
\end{equation}
\begin{equation}
  H(a) = \sqrt{\Omega_{\rm{m}} a^{-3} +
    (1-\Omega_{\rm{m}}-\Omega_{\Lambda}) a^{-2} 
    +\Omega_{\Lambda}}.  
\end{equation}

Although a compact expression using the incomplete beta function has
been shown in Ref.~\citen{bbk}, so far, no analytic solution of
$D_1(a)$ has been presented for $\Omega_{\Lambda}\neq 0$.  Here we
restrict ourselves to the case $\Omega_{\rm{m}}+\Omega_{\Lambda}=1$,
and present an approximate formula in a simple algebraic form.

We adopt a normalization for $D_1(a)$ as
\begin{equation}
    D_1(a) =\frac{5\Omega_{\rm{m}}}{2} H(a) \int_0^a \frac{da'}{\big(a'
      H(a')\big)^3}. 
\end{equation}
Then, in a flat cosmology $\Omega_{\rm{m}}+\Omega_{\Lambda}=1$, our 
formula is
\begin{equation}\label{eq:b4}
  D_1(a)=a\sqrt{1+x}\frac
  {1+1.175x+0.3064x^2+0.005355x^3}
  {1+1.857x+1.021x^2+0.1530x^3}, 
\end{equation}
where, 
\begin{equation}\label{eq:bthree}
  x = \frac{1-\Omega_{\rm{m}}}{\Omega_{\rm{m}}} a^3 . 
\end{equation}
Note that our approximate formula is  exact
when $\Omega_{\rm{m}}=1$. 

The well-known approximation formula for the growth function in Ref.~\citen{carroll}, which was adopted from
Ref.~\citen{lahav},   is
\begin{equation}\label{eq:bcarroll}
  D_1^{\rm{C}} = {\frac{5\Omega_{\rm{m}}}{2}} \frac{1}{
    \Omega_{\rm{m}}^{\frac{4}{7}} - (1-\Omega_{\rm{m}}) +
    \left(1+\frac{\Omega_{\rm{m}}}{2}\right)
    \left(1+\frac{1-\Omega_{\rm{m}}}{70}\right) }  
\end{equation}
for $\Omega_{\rm{m}} + \Omega_{\Lambda}=1$. 
A comparison of the relative error $\Delta E$ at $a=1$  is shown in 
Table~\ref{table:carroll} for $\Omega_{\rm{m}} + \Omega_{\Lambda}=1$. 
Our formula has generally smaller relative error in the range 
$0.2 < \Omega_{\rm{m}} < 1$. 
\begin{table}
  \caption{A comparison of the relative error $\Delta E(\%)$  
 of  Eq.~(\ref{eq:b4}) and Eq.~(\ref{eq:bcarroll}) at $a=1 (z=0)$. }
\label{table:carroll}
\begin{center}
\begin{tabular}{ccccc}
\hline\hline
$\Omega_{\rm{m}}$ & 0.2 & 0.3 &  0.5 & 0.9 \\
\hline
$\Delta E$ of Eq.~(\ref{eq:b4})
&  0.19\% & $<0.01\%$ &  $<0.01\%$ & $<0.01\%$ \\   
$\Delta E$ of Eq.~(\ref{eq:bcarroll}) 
&  0.54\% & \ \ 0.134\% &  \ \ 0.057\% & \ \ 0.019\% \\   
\hline 
\end{tabular}
\end{center}
\end{table}

\end{document}